\documentclass[12pt,preprint]{aastex}

\def\deg{\ifmmode^\circ\else$^\circ$\fi}

\def\mic{~$\mu$m}

\def\mic{$\mu${\rm m}}
\def\lir{$L_{\rm IR}$}

\def\arcs{\ifmmode {''}\else $''$\fi}
\def\arcm{\ifmmode {'}\else $'$\fi}
\def\parcs{\sa=.07em \sb=.03em
     \ifmmode $\rlap{.}$^{\scriptscriptstyle\prime\kern -\sb\prime}$\kern -\sa$
     \else \rlap{.}$^{\scriptscriptstyle\prime\kern -\sb\prime}$\kern -\sa\fi}
\def\parcm{\sa=.08em \sb=.03em
     \ifmmode $\rlap{.}\kern\sa$^{\scriptscriptstyle\prime}$\kern-\sb$
     \else \rlap{.}\kern\sa$^{\scriptscriptstyle\prime}$\kern-\sb\fi}

\def\Lsun{L$_{\odot}$}
\def\lsun{L$_{\odot}$}

\def\spose#1{\hbox to 0pt{#1\hss}}
\def\simlt{\mathrel{\spose{\lower 3pt\hbox{$\mathchar"218$}}
     \raise 2.0pt\hbox{$\mathchar"13C$}}}
\def\simgt{\mathrel{\spose{\lower 3pt\hbox{$\mathchar"218$}}
     \raise 2.0pt\hbox{$\mathchar"13E$}}}
\def\lsim{\rlap{$<$}{\lower 1.0ex\hbox{$\sim$}}}
\def\gsim{\rlap{$>$}{\lower 1.0ex\hbox{$\sim$}}}

\begin{document}

\title{Measuring PAH Emission in Ultradeep {\it Spitzer}$^1$\ IRS$^2$
  Spectroscopy of High Redshift IR Luminous Galaxies}

\altaffiltext{1}{Based on observations obtained with the {\it Spitzer Space Telescope}, which is 
operated by JPL, California Institute of Technology for the National Aeronautics and Space Administration}

\altaffiltext{2}{The IRS is a collaborative venture between Cornell
 University and Ball Aerospace Corporation that was funded by NASA
   through JPL.}

\author{H. I. Teplitz\altaffilmark{3},   
V. Desai\altaffilmark{4}, 
L. Armus\altaffilmark{3}, 
R. Chary\altaffilmark{3}, 
J. A. Marshall\altaffilmark{5}, 
J.W. Colbert\altaffilmark{3}, 
D.T. Frayer\altaffilmark{3}, 
A. Pope\altaffilmark{6},  
A. Blain\altaffilmark{4},  
H. Spoon\altaffilmark{5},   
V. Charmandaris\altaffilmark{7,8},  
D. Scott\altaffilmark{6}  
}

\altaffiltext{3}{Spitzer Science Center, MS 220-6, Caltech, Pasadena, CA 91125.  hit@ipac.caltech.edu}
\altaffiltext{4}{Astronomy Department, Caltech, Pasadena, CA 91125}
\altaffiltext{5}{Astronomy Department, Cornell University, Ithaca, NY  14853}
\altaffiltext{6}{Dept. of Physics \& Astronomy, University of British Columbia, Vancouver, BC V6T1Z1, Canada}
\altaffiltext{7}{Dept. of Physics, University of Crete,  GR-71003 Heraklion, Greece}
\altaffiltext{8}{IESL / Foundataion for Research and Technology-Hellas, GR-71110, Heracklion, Greece, and Chercheur Associ\'e,
Observatoire de Paris, F-75014, Paris, France}

\begin{abstract}
  
  The study of the dominant population of high redshift IR-luminous
  galaxies ($10^{11}$\ - $10^{12}$\ \lsun\ at $1<z<3$), requires
  observation of sources at the $\sim 0.1$\ mJy level in the mid-IR.
  We present the deepest spectra taken to date with the Infrared
  Spectrograph (IRS) on the {\it Spitzer Space Telescope}.  We
  targeted two faint ($f_{24}\sim 0.15$\ mJy) sources in the Southern
  GOODS field at $z=1.09$\ and $z=2.69$.  Spectra of the lower
  redshift target were taken in the observed-frame 8--21 micron range,
  while the spectrum of the higher redshift target covered 21--37
  microns.  We also present the spectra of two secondary sources
  within the slit.  We detect strong PAH emission in all four targets,
  and compare the spectra to those of local galaxies observed by the
  IRS.  The $z=1.09$\ source appears to be a typical, star-formation
  dominated IR-luminous galaxy, while the $z=2.69$\ source is a
  composite source with strong star formation and a prominent AGN.
  The IRAC colors of this source show no evidence of rest-frame
  near-infrared stellar photospheric emission.  We demonstrate that an
  AGN which contributes only a small ($\sim 10$\%) fraction of the
  bolometric luminosity can produce enough hot dust emission to
  overwhelm the near-IR photospheric emission from stars.  Such
  sources would be excluded from photometric surveys which rely on the
  near-IR bump to identify starbursts, leading to an underestimate of
  the star formation rate density.

  \end{abstract}


\section{Introduction}

More than 70\% of present day stars were formed at redshifts $>0.5$\
in galaxies undergoing an infrared luminous phase \citep{Elbaz 2002},
identifiable as LIRGs ($10^{11}<$\lir$< 10^{12}$\ \Lsun) and ULIRGs
(\lir$> 10^{12}$\ \Lsun).  These sources are responsible for the peak
of the faint 24 micron number counts \citep{Papovich 2004, Marleau
  2004}.  LIRGs are particularly important at $z<1.5$, beyond which
ULIRGs may dominate \citep{Lagache 2004, Chary 2004, Perez-Gonzalez
  2005, Le Floch 2005}.  As much as 85\%\ of their luminosity may be
generated by star formation \citep{Bell 2005, Brand 2006}, although
that fraction is still uncertain.

IR luminous galaxies are rare in the present epoch, even though they
may represent a generic phase typical of star formation activity in
distant galaxies.  Locally, 95\%\ of ULIRGs and $\sim 50$\%\ of LIRGs
are driven by mergers \citep[e.g.\ ][]{Bushouse 2002}, which provide the
primary means of triggering the required amounts of star formation and
AGN activity.  At higher redshifts, however, more gas rich systems may
produce stars at high rates under less extreme conditions \citep[e.g.\
][]{Marcillac 2006,Sajina 2006}.  This evolution is seen in the
optical morphologies of $z\sim 1$\ LIRGs \citep{Bell 2005}, which
span a range of Hubble types, from ellipticals to irregulars.

High redshift IR luminous galaxies are easily detected by {\it Spitzer
  Space Telescope}\ imaging surveys such as the Great Observatories
Origins Deep Survey \citep[GOODS;][]{Giavalisco 2004, Dickinson 2004}
and the SIRTF Wide-Area Infrared Extragalactic Survey
\citep[SWIRE;][]{Lonsdale 2003}.  These sources have been beyond the
reach of most large {\it Spitzer}\ spectroscopy programs of high
redshift sources, which have been limited to $f_{24}>0.7$\ mJy and IRS
exposures of $\simlt 1$\ hour \citep{Houck 2005, Yan 2005, Lutz 2005},
and typically sample only the most extreme ULIRGs ($>10^{13}$\ \lsun).

MIR spectra of starbursts are dominated by emission from polycyclic
aromatic hydrocarbons (PAHs) at 6.2, 7.7, 8.6, 11.3, 12.7, and 17 \mic\
and broad ($\sim 2$\ \mic) silicate absorption at 9.7 and 18 \mic\
\citep[][ and the references therein]{Genzel and Cesarsky 2000}. In
local sources, the presence of PAH features, combined with the shape
of the continuum, can be used to determine the nature of the emission
(AGN vs. star formation) that dominates the bolometric luminosity of a
source \citep[e.g.][]{Laurent 2000, Peeters 2004, Armus 2006, Spoon 2007}.  The
reliability of these same diagnostics at high redshift requires
further demonstration.

LIRGs at $z\sim 1$\ and ULIRGs at $z\sim 2$\ are likely to have flux
densities on the order of 0.1 mJy at $\sim 20$\ \mic, based on the
bolometric corrections of \cite{Chary and Elbaz 2001}.  In principle,
such sources should be observable with the {\it Spitzer} Infrared
Spectrograph \citep[IRS;][]{Houck 2004}\ in several hours, assuming
that the noise characteristics do not change at faint fluxes and the
signal-to-noise ratio scales with the square root of observing time.
In order to push the limits of MIR spectroscopy with {\it Spitzer}, we
targeted two faint (0.15 mJy) galaxies at $z \ge 1$, using 32 hours of
Director's discretionary time.  In this paper, we present the results
of those observations, demonstrating for the first time the capability
of the IRS to detect continuum emission at such faint flux levels.
Throughout, we assume a $\Lambda$-dominated flat universe, with
$H_0=71$\ km s$^{-1}$\ Mpc$^{-1}$, $\Omega_{\Lambda}=0.73,$\ and
$\Omega_{m}=0.27$.

\section{Target Selection}

We selected targets from the southern GOODS field.  GOODS, by virtue
of being the target of the deepest {\it Spitzer, Hubble, Chandra}, and
ground-based imaging and spectroscopy provides the richest, most
homogeneous dataset yet compiled to understand the formation and
evolution of galaxies and AGN.  While the Northern field offers
certain advantages, the Southern field was better positioned for the
available {\it Spitzer}\ scheduling window.  Amongst the unique
components of the Southern field are: the {\it Hubble}\ Ultra-Deep
Field \citep[UDF;][]{Beckwith 2006}, which provides high resolution
ACS imaging; extensive spectroscopic redshifts \citep{Le Fevre 2004};
ultradeep 24 \mic\ \citep{Chary 2006}\ and shallow 16 \mic\ photometry
\citep{Teplitz 2005}; and ultradeep (1 Ms) {\it Chandra}\ 2--10 keV
X-ray observations which provide the most complete census of AGN
activity \citep{Giacconi 2002}.

The goal of the observations was to examine relatively faint
IR-luminous galaxies at high redshift. We estimate the IR luminosity
from the observed 24 \mic\ flux.  The MIR luminosity of local galaxies
in the IRAS bright galaxy sample \citep{Soifer 1987}\ has been found
to correlate strongly with their far-infrared luminosity, which is
dominated by large, cool dust grains \citep{Chary and Elbaz 2001}.
This correlation has been applied to develop a library of model
templates of the mid- and far-infrared SEDs of galaxies.  The library
consists of template SEDs across a range of luminosities, which can be
redshifted to predict the MIR flux of a source with given luminosity
at a redshift of interest.  For each source in UDF, we select the
template for which the library predicts the closest 24 \mic\ flux
density at the appropriate redshift to apply a bolometric correction;
we do not use the shorter wavelength IRAC measurements.  The
corrections based on the \cite{Chary and Elbaz 2001}\ and \cite{Dale
  2002}\ template are used to derive an infrared luminosity
(\lir$=8-1000$\ \mic).  The use of $f_{24}$\ to predict \lir\ using these
templates is supported by the observation of sub-millimeter galaxies (SMGs)
in GOODS-North by \cite{Pope 2006}.

We chose two sources in the UDF, one with a {\it Chandra}\ detection
and one without.  We chose each source to have a ratio (or upper
limit) of X-ray to infrared luminosity of $L_X$/\lir $< 1.5 \times
10^{-3}$, consistent with many SMGs \citep{Alexander 2005}.  We
avoided sources that are only marginally resolved in {\it HST}\ imaging, which might
indicate strong AGN contributions, and selected instead from those
with fully resolved morphologies.  A final criterion was the
spectroscopically-confirmed redshift, which we selected to place the
strongest PAH features in the center of one of the IRS passbands.  By
selecting one source at $z\sim 1$\ and one at $z\sim 2.5$, we were
able to obtain spectra covering the 6.2, 7.7 and 8.6 \mic\ PAH
features in just one or two slits.

\begin{figure}[t!]
\plotone{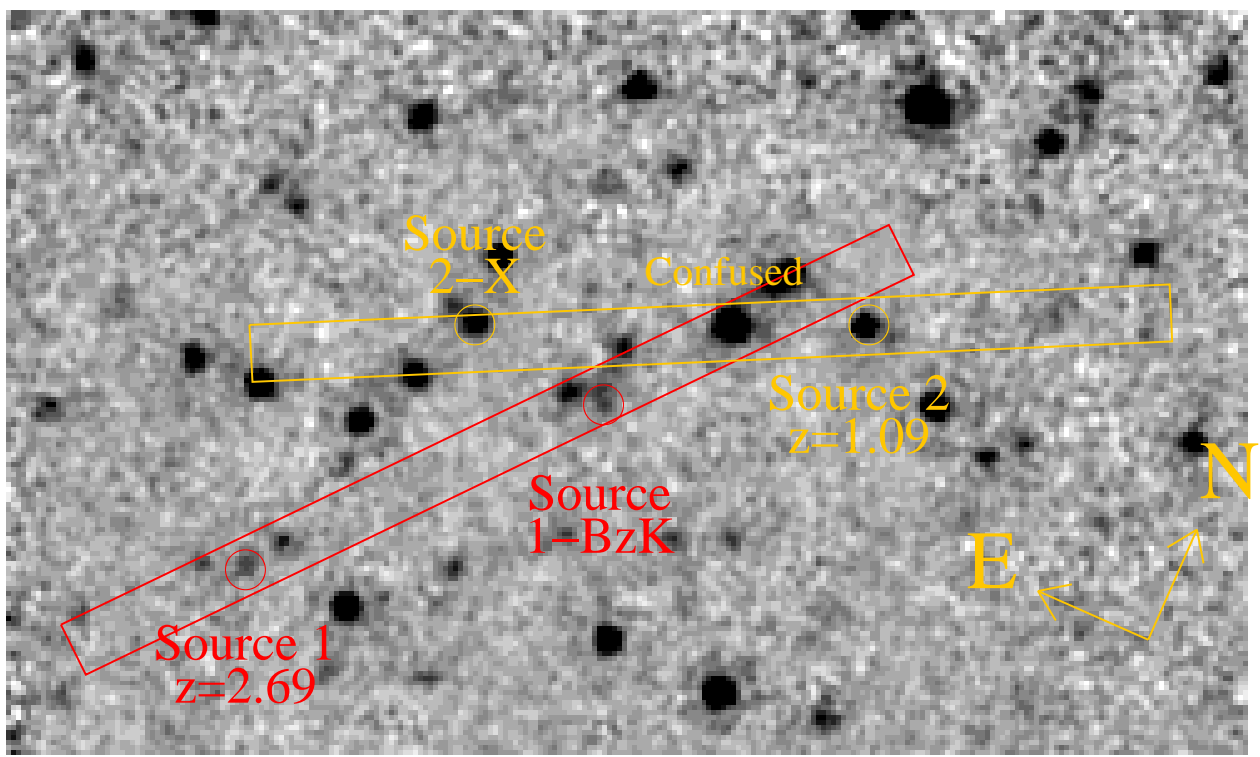}
\caption{\label{fig: overlay} The orientation of the LL-1 (red) and LL-2 (yellow) slits overlayed
on the 16 \mic\ image of the UDF. The image size is $4^{\prime}\times 2.5^{\prime}$.  The primary sources are indicated
by their redshift, as are other sources in the slits.}
\end{figure}

\begin{figure}[t!]
\plotone{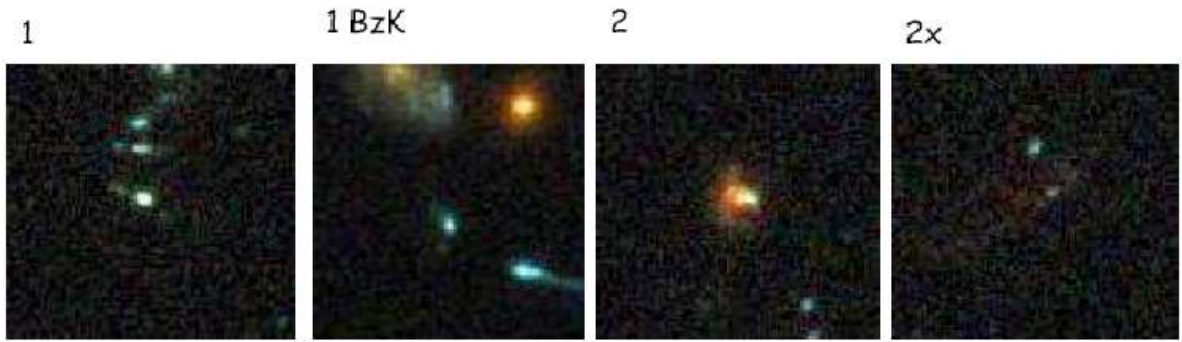}
\caption{\label{fig: stamps} Three-color ACS images of the four sources of interest.  
  The images are the combination of the {\it Viz}\ filters in the
  GOODS data.  Images were obtained from the MAST high level science
  products.  The images are 5 arcseconds on each side; north is up in
  the images, and east is to the left.}
\end{figure}

To meet these criteria, we selected a $z=2.69$\ {\it Chandra}-detected
source with 24 \mic\ flux density of $\sim 0.13$\ mJy, hereafter
Source 1, and a z=1.09 object with a 16 \mic\ flux density of $\sim
0.15$\ mJy and no X-ray detection, hereafter Source 2.  The infrared
luminosity of these sources is approximately $10^{13}$\ and $3\times
10^{11}$, respectively, based on the \cite{Chary and Elbaz 2001}\
templates.  The {\it Chandra}\ detection of Source 1 is only in the
soft-band (0.5-2 keV), which does not immediately rule out
star-formation as the origin of the X-ray emission. Given the
experimental nature of the observations, we were given more than usual
latitude to arrange the precise scheduling of the observations.  Thus,
we were able to choose observation dates which were favorable for
including additional sources in the slit.  Figure \ref{fig: overlay}\
shows the position of the slit overlayed on the 16 \mic\ map of the
field.  Two of the secondary objects in the slits are of great
interest, and are indicated in the figure: (1) an X-ray source
detected by {\it Chandra}\ in the soft-band only, but lacking an
optical redshift, referred to hereafter as Source 2-x; and (2) a
source lacking an optical redshift but selected to be at high redshift
by the {\it BzK}\ selection technique \citep{Daddi 2004}, referred to
hereafter as Source 1-BzK.  Figure \ref{fig: stamps}\ shows the ACS
images of each source.  Table \ref{Tab: phot}\ gives the photometry
for the four sources of interest.  Both LL slits also include a
somewhat brighter object ($f_{16}\sim 0.25$\ mJy) which appears to be
a single source in the 16 \mic\ map, but IRAC imaging reveals it to be
a blend of three sources.  The slits were oriented to include this
source in case the sensitivity proved worsen than predicted and the
primary targets were not detected.  Given its blended nature, it is
not discussed further in the present paper.

The 16 \mic\ map shows that Sources 1 and 1-BzK have close, but
clearly separated, companions of comparable brightness.  They lie
$\sim 8$\arcsec\ from the targets, which is less than two pixels in
LL-1 (5.1\arcsec\ pixels).  Although we do not know that the
companions lie at different redshifts than the targets, we have no
evidence that they do; we expect that they will not affect the
detection of emission features, though they will make the measurement
of equivalent widths less certain.  The companion sources are of
comparable brightness to the targets, so, if half of the companion
light were to fall into the extraction aperture, the continuum would
be overestimated by about 50\%.
\label{sec: targets}

\section{Observations and Data Reduction}

We chose integration times that would likely achieve signal to noise
ratios of $2-4$\ in the continuum, so that the results could be
evaluated even if no emission or absorption features were detected.
We obtained the observations using a different observing mode ({\it Staring}\ or
{\it Mapping}) for each object, in order to compare the effectiveness of the
modes.

For the $z=1.09$\ target, Source 2, we obtained a spectrum using the second
order of the Long-Low module (LL-2, $14-21$\ \mic).  The spectrum was
taken in {\it Staring Mode}, in which the target is placed at two
``nod'' positions within the slit.  The observation was broken into
two separate Astronomical Observing Requests (AORs), each of which
contained 4.5 hours of on-source integration divided between the two
nod positions.  At each nod position we took 70 spectra of 120 sec
ramp duration.  The total on-source integration time was 9.4 hours,
and the total AOR time with overhead was 11.9 hours.  We also observed
this target in the short-wavelength low-resolution slit, in the first
order (SL-1), in a single AOR for three hours on source in {\it
  Staring Mode}.

For the higher redshift target, Source 1, we used the first order of the
Long-Low module (LL-1, $21-37$\ \mic).  These observations were
obtained in {\it Mapping Mode}, wherein the target is placed at
multiple positions along the slit; we selected six map positions.  The
observations were broken into three AORs, each with 20 spectra of 120
sec ramp duration at each of the six map positions.  The total
on-source integration time of 12 hours, and the total AOR time with
overhead was 15.4 hours.  The observations are summarized in Table
\ref{Tab: obs}.

Observations were obtained on 06 August 2005 (SL-1), 10 August 2005
(LL-2), and 12 September 2005 (LL-1).  The LL-2 AORs were executed
consecutively.  Two of the LL-1 AORs were consecutive, but there was a gap
of more than 12 hours between the second and third observation.  We
observe no degradation in signal to noise resulting from the gap.  All
observations were scheduled immediately after the ``skydark''
calibrations in order to maximize the sensitivity by reducing the
change of latent images from a preceding bright target.  

To guard against the unlikely event of large deviations in telescope
pointing, we obtained high accuracy peak-ups in the blue filter, using
a star selected from the 2MASS catalog.  During observation planning,
we examined the MIPS 24 \mic\ image of the star prior to submitting
the AORs to ensure that its flux was accurately predicted and that it
had no bright neighbors.  We inspected the PU data, to check that the
centroid was reasonable and that the pointing was as expected.  No
problems were identified.

\subsection{Data Reduction}

The Basic Calibrated Data (BCD) were produced by the S13 pipeline at
the {\it Spitzer} Science Center (SSC), which includes ramp fitting,
dark sky subtraction, droop correction, linearity correction, flat
fielding, and wavelength calibration (see the IRS Pipeline
Handbook\footnote{http://ssc.spitzer.caltech.edu/irs/dh/}).  Further
processing of the two dimensional dispersed frames is required before
spectral extraction.  In addition to the pipeline processing, we
performed the following reductions: latent charge removal, rogue pixel
interpolation, and residual sky subtraction.

A small fraction (1-2\%) of charge on the detector persists between
frames despite the resetting of the detector that occurs prior to
every integration. This latent charge decays slowly over time and is
removed by the annealing process. In the case of very faint sources,
the source of latent charge is the zodiacal background.  Over the
course of a six hour AOR, this charge can build up to a significant
level.  We removed the latent signal by fitting the slope of the
background with time.  We measured the median background independently
for each row of each frame, fitted the slope of these median values,
and then subtracted this fit row-by-row from each frame. The
background in BCD images is the residual after ``skydark'' subtraction
in 3D.  In the deep observations, the residual sky level was $\sim
50$\ electrons per second, and the latent charge built up to $\sim 5$\ 
electrons per second by the end of the AOR.
  
Unstable, or ``rogue'', pixels are those which are usable in some AORs
and not in others, depending on the recent history of the detector.  A
mask of known rogue pixels is provided by the SSC, and we identified
further suspect pixels from the data themselves.  We used the {\sc
  IRSCLEAN}\ program provided by SSC to find rogue pixels in the 2D
data, using its default settings.  We further searched for pixels with
abnormally high variance ($> 10\sigma$) with time.  All pixels identified as rogue
were interpolated over using {\sc IRSCLEAN}.
  
Once the BCD data were ``cleaned'', we created residual sky images
from the data.  In staring mode, we used the other nod position, and
in mapping mode we used the other five map positions.  In both cases,
we used a resistant mean (removing outliers before calculating the
mean) with time to calculate the sky value in each pixel.

The individual reduced frames were coadded to produce final 2D spectra
at each nod or map position.  One dimensional spectra were extracted
using the {\sc SPICE}\ software provided by the SSC.  To minimize the
noise contribution from the background, we used narrower extraction
windows than the default.  We used a window which expanded with
wavelength and had a width of $\sim 2$\ pixels at the blue end of each
order.  The ``slit losses'' introduced by the narrow window were
estimated by extracting the spectrum of the standard star HR 7341
(taken in the same campaign) using both the standard and the new
width.  The correction factor varied with wavelength by $\sim 10$\%\ 
and had a mean value of 1.4.  The spectra of HR 7341 were taken at the
standard {\it Staring Mode}\ nod positions, so the {\it Mapping Mode}\ 
spectra may not be perfectly corrected (at the $<5$\%\ level), due to
flat field variation or pixel undersampling.

\subsection{Noise Measurement}

The signal to noise ratio (SNR) in the IRS spectra was measured using
the variance of each pixel from frame to frame.  We measured the
variance in the sky frame for each pixel, scaled by e.g.  $\sqrt{2}$\
for the nodded frames, and created two dimensional noise frames for
input into {\sc SPICE}.  The uncertainties were then propagated by the
extraction process.  These uncertainties were checked against the
standard deviation in the continuum of the 1D spectra at, for example,
$\sim 24$\ \mic\ in LL-1.

The SNR at 16 \mic\ in LL-2 and 24 \mic\ in LL-1 is generally
consistent with the predictions of the SSC website ({\sc SPEC-PET}).
The predictions do not include noise introduced by the
pipeline processing or spectral extraction.  They also assume the
spectra are smoothed to a resolution of $R=50$, which we have not
done.  A 12 hour integration in LL-1 is expected to
yield a $3\sigma$\ uncertainty of 0.12 mJy at 24 \mic\ with spectral
resolution smoothed to $R=50$.  Our measured $3\sigma$\ uncertainty is
$\sim 0.13$\ mJy.  {\sc SPEC-PET}\ suggests a 9 hour integration in
LL-2 will yield a $3\sigma$\ uncertainty of 0.08 mJy at 16 \mic\ and
$R=50$.  Our measured uncertainty is $\sim 0.12$\ mJy.  While both
spectra achieve sensitivity close to the estimates, the
{\it Mapping Mode}\ observation gives noticeably better results.  Many
factors may contribute to the difference between the modes, including
better sky estimation in {\it Mapping Mode}, as well as an averaging
over low level variation in the flat field.  

The SNR varies strongly with wavelength, with noise rising sharply at
the red end of each order.  The shape of the error array measured in
our data agrees well with the predictions of {\sc SPEC-PET},
independent of the overall level.

\section{Results}

\begin{figure}[t*]
\includegraphics[scale=1.0]{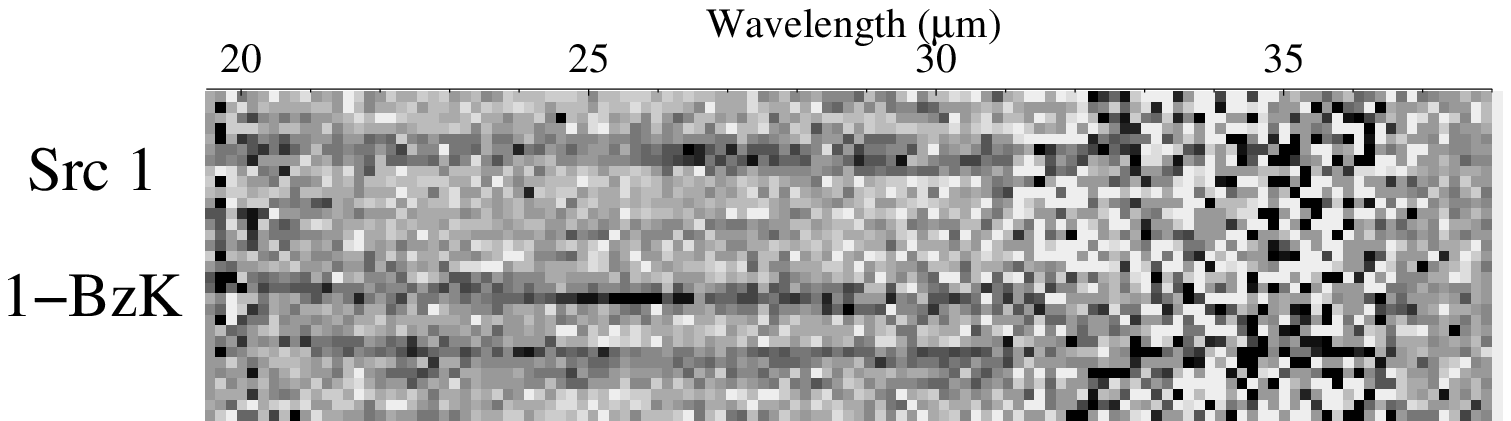}
\includegraphics[scale=1.0]{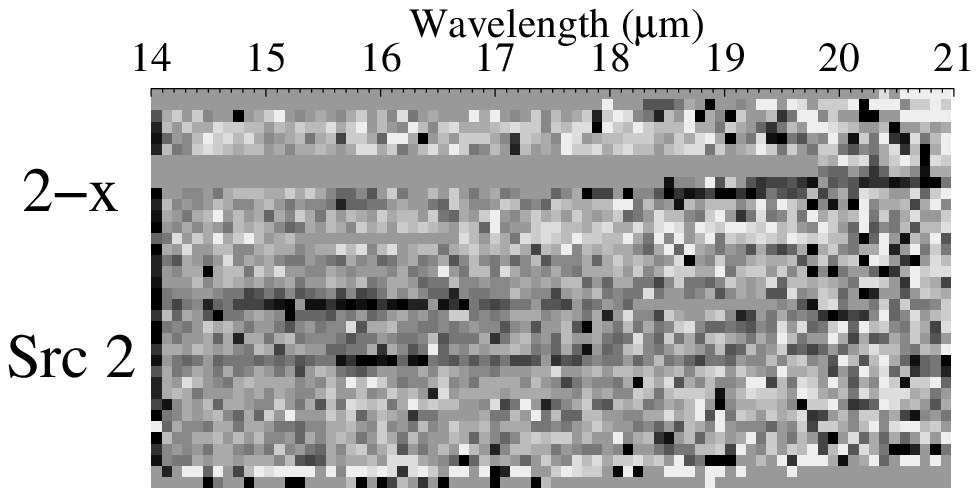}
\caption{\label{fig: image} The 2D dispersed frame image of the IRS spectra of faint sources.  
  The top panel shows the LL-1 spectrum of Source 1 and Source 1-BzK,
  in one of the six map positions.  The bottom panel shows the LL-2
  spectrum of Source 2 and Source 2-x in one of the two nod positions.
  Many of the pixels just above Source 2-x have been masked. In each
  case, only the illuminated portion of the relevant order is shown.
  The approximate wavelength scale is indicated. Positive signal is
  black on white.  }
\end{figure}

\begin{figure}[t*]
\plotone{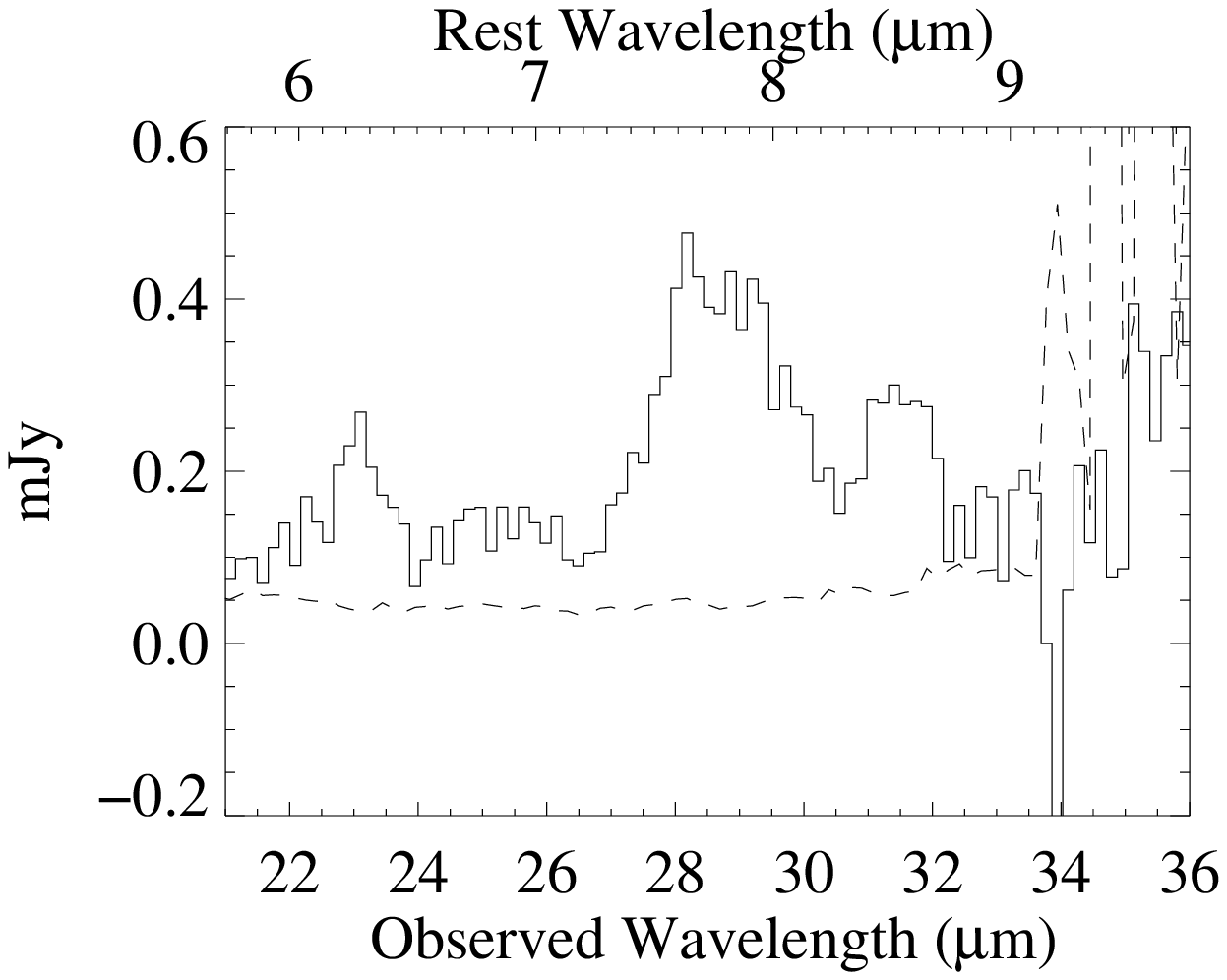}
\caption{\label{fig: spectra 1} 
  The spectrum of  Source 1 at $z=2.69$\ in LL-1.  The $1\sigma$\ error arrays are
shown (dashed lines). }
\end{figure}

\begin{figure}[t*]
\plotone{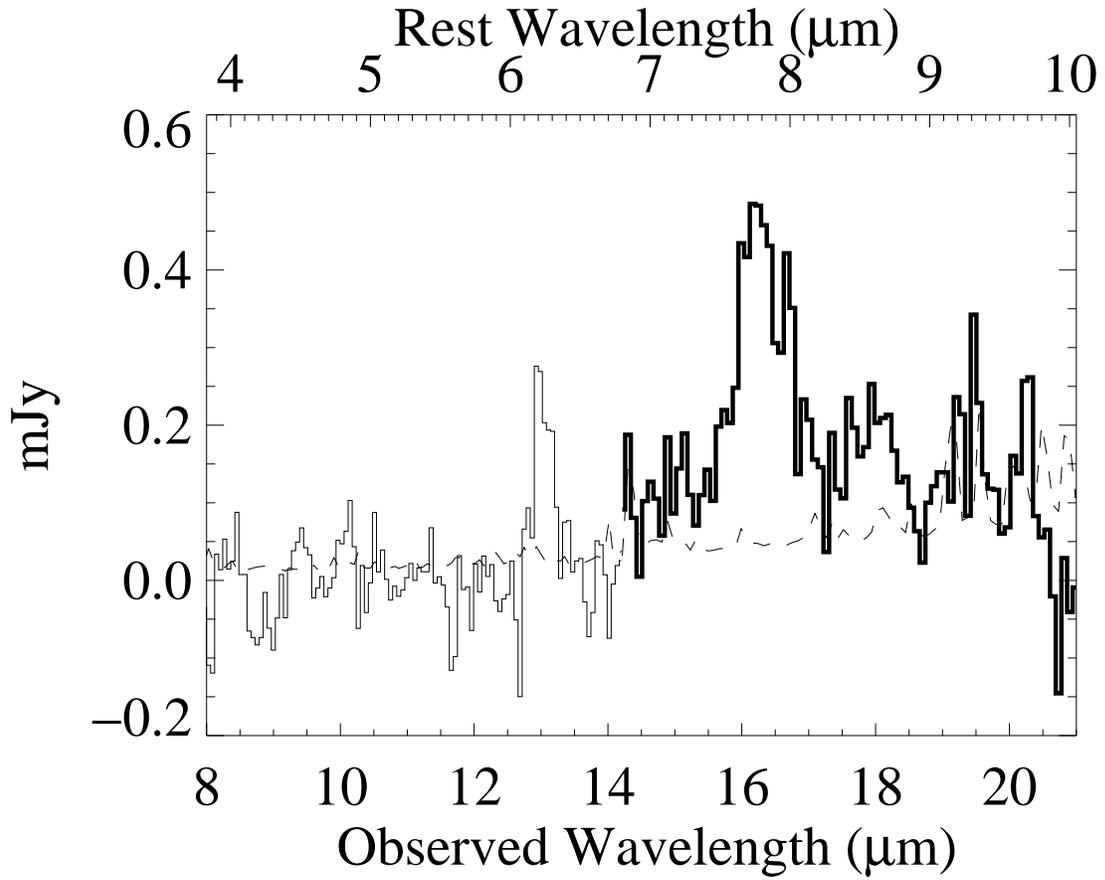}
\caption{\label{fig: spectra 2} 
  The spectrum of Source 2 at $z=1.09$\ in SL-1 (thin line) and LL-2
  (thick line).  The $1\sigma$\ error arrays are shown (dashed
  lines).}
\end{figure}

\begin{figure}[t*]
\includegraphics[scale=1.0]{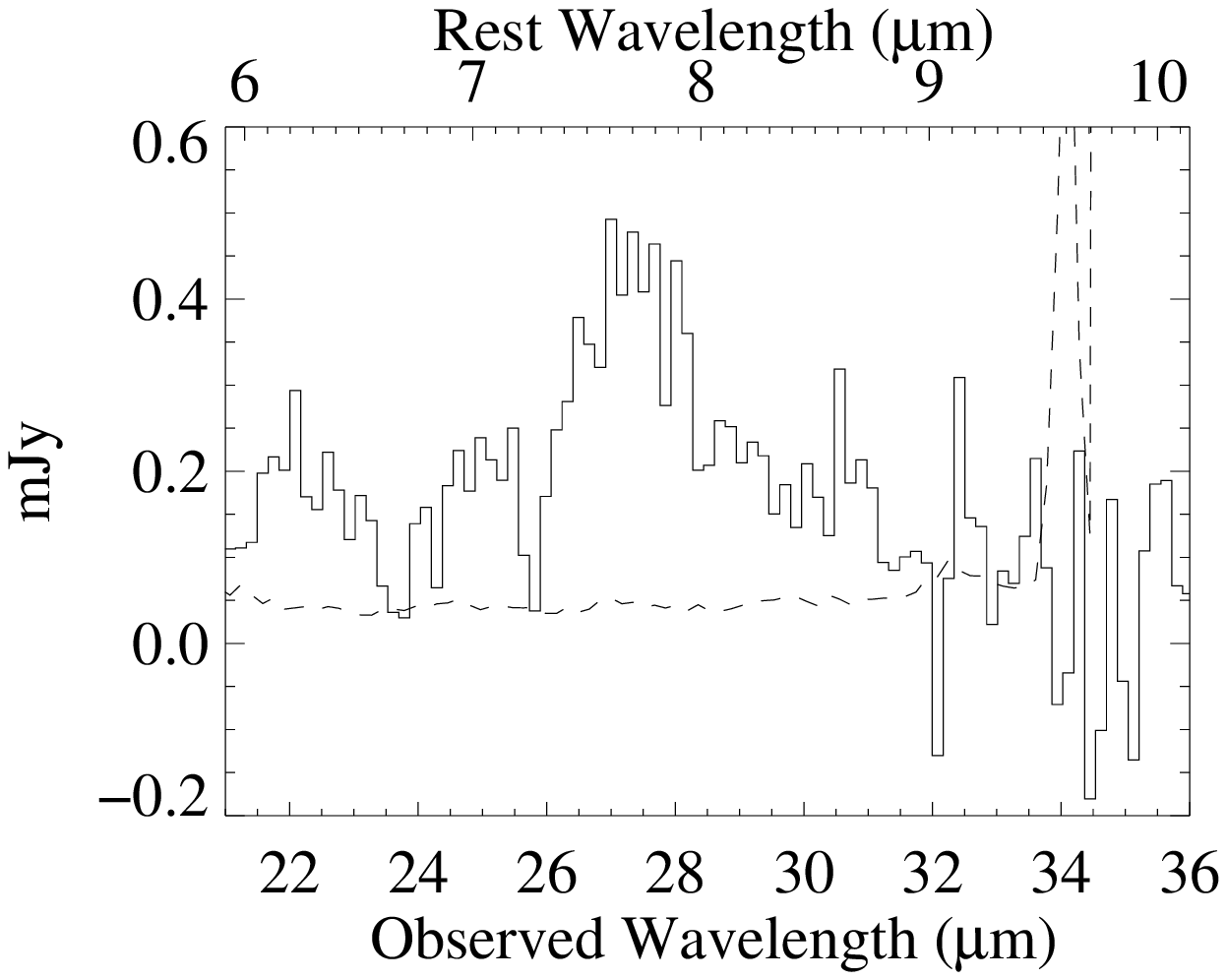}
\caption{\label{fig: spectra 3} 
  The spectrum of Source 1-BzK at $z=2.55$\ in LL-1.  The $1\sigma$\ error arrays are shown
  (dashed lines).}
\end{figure}

\begin{figure}[t*]
\includegraphics[scale=1.0]{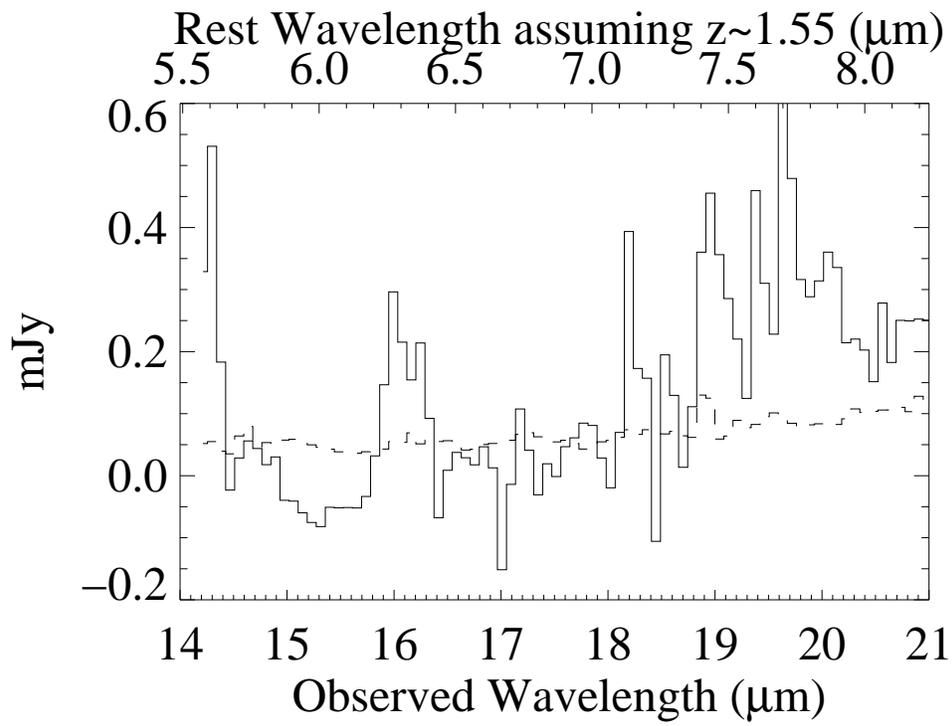}
\caption{\label{fig: spectra 4} The spectrum of Source 2-x in LL-2,
  likely at $z\sim 1.55$.  The $1\sigma$\ error arrays are shown
  (dashed lines). }
\end{figure}

Figure \ref{fig: image}\ shows the 2D spectra of the sources.  Both
primary sources and secondary targets discussed above (Source 1-BzK
and Source 2-x) are detected.
Figures \ref{fig: spectra 1}--\ref{fig: spectra 4}\ show the
extracted spectra of the primary and secondary targets. The continuum
level is in good agreement with the estimate from the broad-band
photometry at 16 and 24 \mic.  Broad emission features, identified as
polycyclic aromatic hydrocarbon (PAH) emission, are detected at rest
wavelengths of 6.2, 7.7, and 8.6 \mic, along with the continuum at the
blue end of LL-2 and LL-1 (except in Source 2-x).

The PAH features in Source 1-BzK confirm that it is at high redshift,
indicating $z=2.55 \pm 0.03$.  The uncertainty here is based upon the
narrower 6.2 \mic\ PAH emission feature, and excludes the broader 7.7
\mic\ feature.  This redshift is consistent with the expectation from
the optical photometry.  The other secondary source, Source 2-x, shows
a single definite emission line at an observed wavelength of
$\lambda_{obs}\sim16$\ \mic, and possibly broad enhanced emission at
$\lambda_{obs}\sim 19$\ \mic.  If the isolated line were identified
with the 6.2 \mic\ PAH feature and the broad emission with the 7.7
\mic\ PAH feature, the object would lie at $z\sim 1.55$,.
Identification of the isolated emission line as the 7.7 \mic\ PAH
feature (indicated $z\sim 1.1$) is less likely, as that would center
the broad emission at rest-frame 9.5 \mic, which might only be
explained by strong silicate emission, which is unusual in a source
with PAH emission.

Many infrared luminous sources show deep absorption troughs at
rest-frame $\sim 10$\ \mic\ from silicates.  This rest-frame
wavelength is at the red end of the order for all target spectra and
thus the presence of the absorption is difficult to establish.  

We present estimates of the equivalent width (EQW) of the PAH emission
lines in each object in Table \ref{Tab: EW}.  The continuum was
measured on both sides of the line, when detected, and assumed to have
negligible slope across the line. The continuum in all four objects is
similar, showing shallow or no slope across the modest restframe
wavelength range covered by each order.

\section{Discussion}

Sources 1, 2, and 1-BzK all show strong PAH emission.  Source 2-x has
emission that may arise from PAHs as well.  Figure \ref{fig:
  starburst}\ compares the spectra of each source with the starburst
template spectrum of \cite{Brandl 2006}.  The PAH features in all four
sources are broadly similar to the template.  Local starbursts and
star-forming ULIRGs typically have rest-frame equivalent widths
between $\sim 0.5$\ and $\sim 0.8$\ \mic\ for both the 6.2 and 7.7
\mic\ features, and equivalent width ratios between these features of
around unity \citep{Brandl 2006, Armus 2006}.  Our UDF sources show
slightly lower rest-frame EQW(6.2), and the ratio of EQW(7.7)/EQW(6.2)
appears larger by a factor of about two in the $z>2$\ sources and a
factor of about 1.5 for the $z=1.09$\ source.  This difference is
largely attributable to the low continuum near the 7.7 \mic\ feature
in the high redshift objects.  The continuum is detected only at 3--4
$\sigma$\ confidence, so this result is hard to quantify.  The
presence of close companions to the $z>2$\ sources also contributes to
the lower EQW (see section \ref{sec: targets}).

At low redshift, PAH emission is usually associated with star
formation. Sources with strong AGN generally have very weak or absent PAH 
emission.  The strength of the PAH emission can
thus be used to estimate the relative contribution of star formation
to the bolometric luminosity \citep[e.g.][]{Laurent 2000, Peeters
  2004}, so these spectra are suggestive of starburst sources.  It is
not surprising that Source 1-BzK or Source 2, a $z=1.09$\ LIRG, show
prominent PAH emission.  However, one might have expected Source 1,
which is detected in X-rays, to have weaker emission.

\begin{figure}[t*]
\plotone{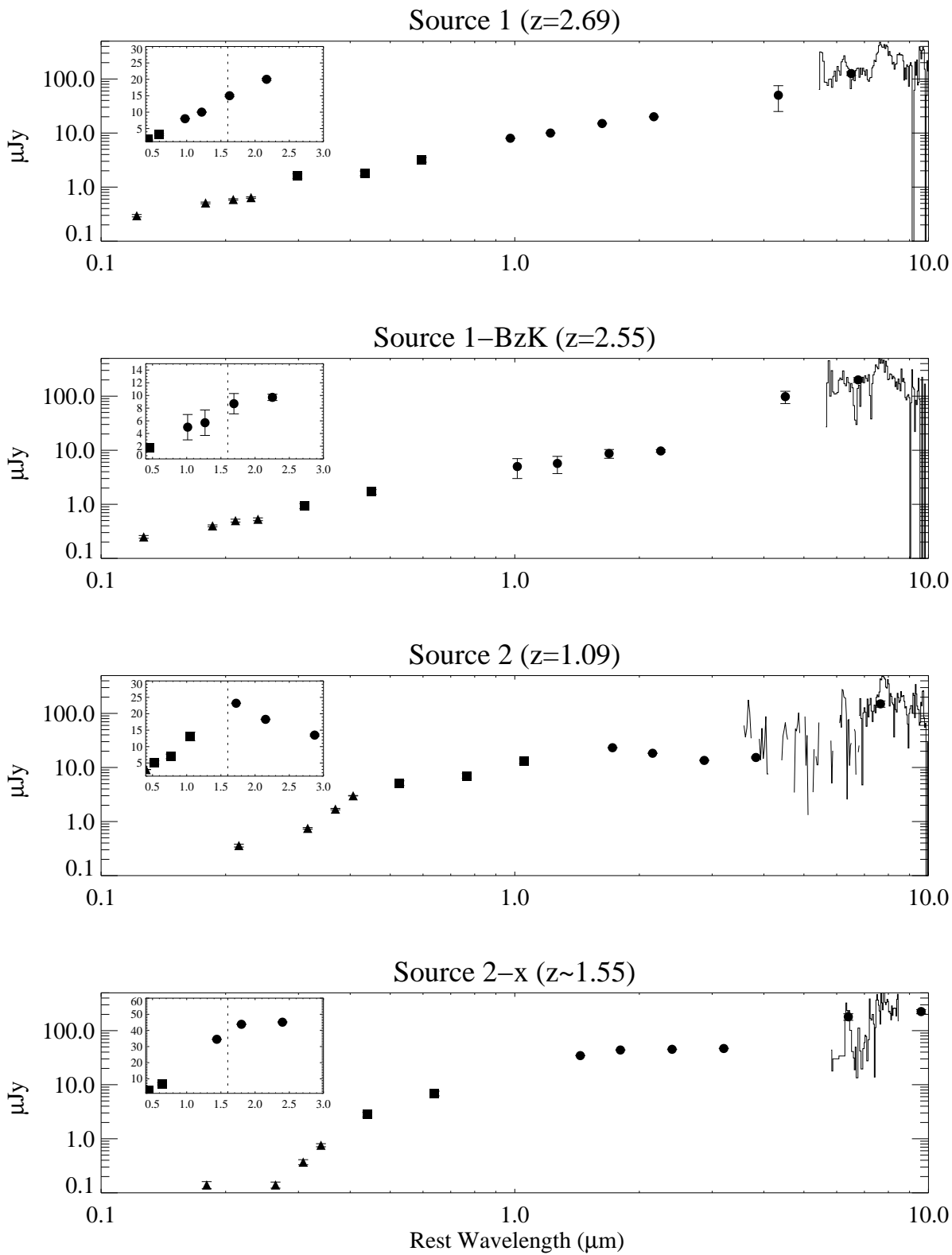}
\caption{\label{fig: SED} The spectral energy distributions (SEDs) of
  the four UDF sources, using the the IRS spectra together with the
  {\it HST}\ (triangles), ground-based NIR (squares), and {\it
    Spitzer}\ (circles) photometry.  Inset plots show the wavelength
  region around 1.6 \mic\ (dotted line) on a linear scale. }
\end{figure}

The wavelength coverage of the present observations is quite limited.
We can extend the SED by considering additional observations of the
GOODS fields: {\it Spitzer} IRAC data at 3.6, 4.5, 5.8, and 8 \mic;
the 16 \mic\ IRS photometry (Teplitz et al.\ 2007, in prep.); and the
HST/ACS and ground-based NIR imaging \citep{Giavalisco 2004}.  These
extended SEDs are shown in Figures \ref{fig: SED}.  The figure inset
shows the wavelength region around the $\sim 1.6$\ \mic\ opacity
minimum in the stellar atmospheres of AGB stars.  This photospheric
emission ``bump'' is commonly observed in galaxies with large evolved
stellar populations and is often used to identify potential massive
starbursts \citep[e.g.  ][]{Lonsdale 2005}, as it is sensitive to a
{\it lack}\ of hot dust emission, which would drown out the ``bump.''
However, the feature may also be prominent in relatively modest
star-forming galaxies if they lack significant dust emission.

The extended SED shows that Source 2, at $z=1.09$, appears indeed to
be a typical, stellar-dominated LIRG, as predicted by its other
properties.  A clear emission peak is seen in the IRAC photometry at
1.6 \mic\ in the rest-frame.  

Source 1 at $z=2.69$, on the other hand, does not show this ``bump.''
In fact, its SED appears relatively featureless at all wavelengths
shortward of 6.2 \mic, extending into the NIR and optical.  This
featureless SED is consistent with the expectation for an X-ray source
at high redshift.  Source 1, therefore, is likely to be a composite
object containing an AGN which dominates the emission in the near-IR,
swamping the 1.6 \mic\ bump, together with significant ongoing star
formation which is responsible for the PAH emission in the rest-frame
MIR.  We also find that source 1-BzK has a MIR SED similar to that of
Source 1, lacking a prominent 1.6 \mic\ bump.

If Source 1 is, indeed, a composite object, then our estimate of its
luminosity, $\sim 10^{13}$\ \lsun\ based on a starburst template, may
be too high.  We can refine this estimate by scaling the luminosity of
a local composite source.  UGC 5101 is a local
ULIRG that shows significant PAH emission, but also has optical line ratios
consistent with an AGN which contributes less than 10\% of the
bolometric luminosity, but deep silicate absorption leaves open the
possibility of a larger contribution \citep{Armus 2004}. Normalizing
the SED of UGC 5101 at 24 \mic\ to Source 1, we estimate the bolometric luminosity
of Source 1 to be $4\times 10^{12}$\ \lsun, a factor of about 2.5 less than 
using a pure starburst template.  If we attribute all of the flux in
Source 1 at restframe 5 \mic\ to an AGN, and scale a representative
AGN template \citep[05189-2524;][]{Armus 2006}\ to that flux, we find
a contribution of $1.6 \times 10^{12}$\ \lsun\ from the AGN to \lir.

\label{sec: newLIR}

Using the UGC 5101 template suggests ratio of $L_X/$\lir$\sim 0.004$\ 
in Source 1, consistent with the ratio observed for radio-detected
SMGs with significant AGN contribution \citep{Alexander 2005}.  The
{\it Chandra}\ non-detection of Source 1-BzK and Source 2 sets limits
on their $L_X/$\lir\ ratios of about 0.001, which does not rule out
the presence of a weak AGN in those objects.  Source 2-x has a ratio
of $\sim 0.007$\ (similar to that of an AGN-dominated SMG) if it is at
the adopted redshift.

\subsection{The 1.6 \mic\ Bump vs. PAH emission}

\begin{figure}[t*]
\plotone{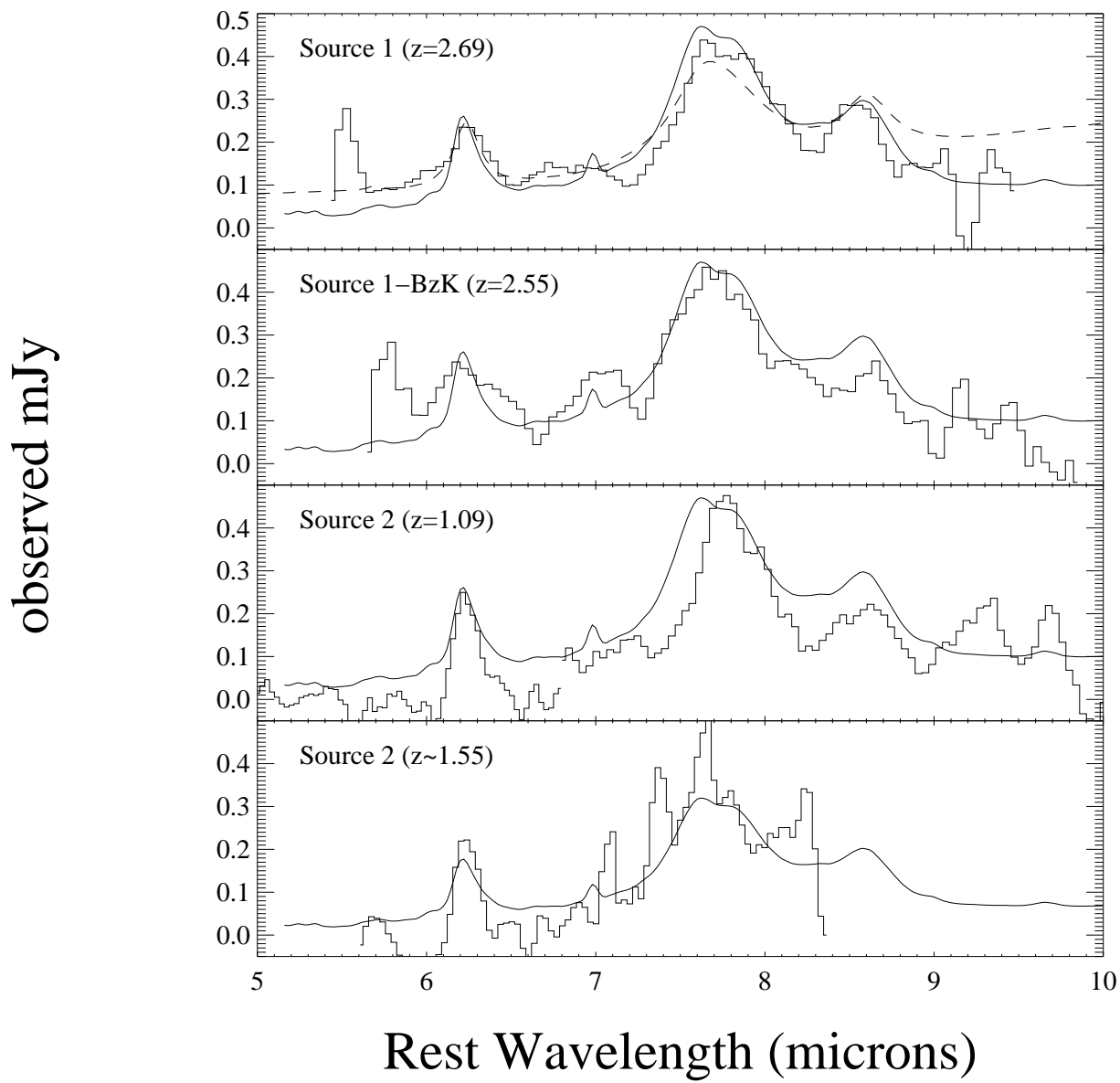}
\caption{\label{fig: starburst}
The spectra of each of the four UDF sources compared to the starburst 
template of \cite{Brandl 2006}\ (smooth solid line).  The UDF source spectra have been smoothed
by 2 wavelength elements.  The template spectrum has been normalized to match the peak of
the PAH emission features.  The spectrum of Source 1 is also compared to the model of a 
starburst-AGN composite (dashed line; see Section \ref{sec: bump-plot}), with a 10\% AGN contribution.}
\end{figure}

Source 1 likely includes a significant AGN contribution, combined with 
strong PAH features but lacks a 1.6 \mic\ bump.  To quantify the sensitivity
of the bump to the strength of an AGN contribution, we construct a
simple model of composite objects by mixing the SEDs of a local
starburst (NGC 7714) and a pure QSO (PG 0804+761), both of which are
first normalized to have unit 1--1000 \mic\ luminosity. We find that
even a weak (10\%\ of the mixture) AGN may produce enough hot dust
emission in the NIR to hide the 1.6 \mic\ bump, while leaving the PAH 
features easily detectable.  This hot dust continuum does, however, noticeably
reduce the EQW of the PAH emission.  We plot a 10\% AGN composite in comparison
to Source 1 in Figure \ref{fig: starburst}.

\label{sec: bump-plot}

We can compare the PAH emission with the stellar bump in a variety of other 
objects.  We can quantify the strength of the 1.6 \mic\ bump using the
ratio of the rest-frame near-IR $J/K$\ to $K/H$\ slopes:

\begin{equation}
B = {{[f_{\nu}(J)/f_{\nu}(K)]}\over{[f_{\nu}(K)/f_{\nu}(H)]}}
\end{equation}

\begin{figure}[t*]
\plotone{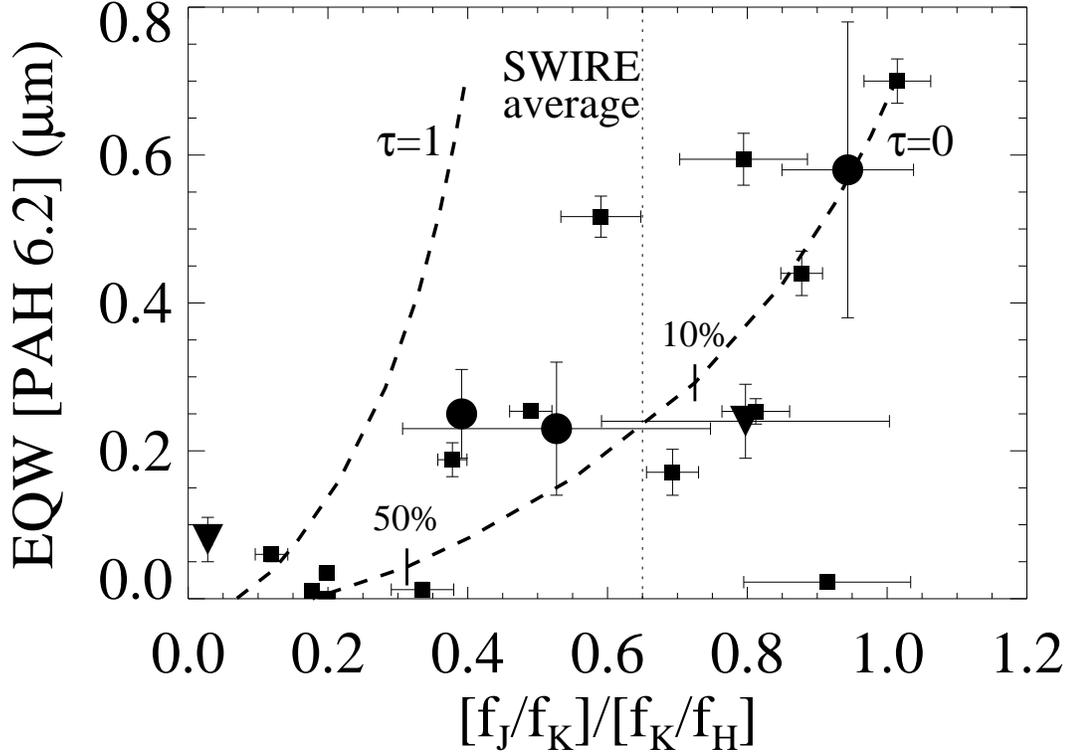}
\caption{\label{fig: 6.2-bump} The strength of the 1.6 \mic\ bump,
  parameterized as the ratio of NIR slopes, vs. the rest-frame EQW of
  the 6.2 \mic\ PAH feature (in microns).  We compare UDF sources
  (circles; this work), FLS sources \citep[triangles;][]{Yan 2005},
  and local ULIRGS \citep[squares;][]{Armus 2006}. For local ULIRGs,
  we have used the total NIR light to measure the bump, rather than an
  aperture matched to the size of the IRS slit.  The dotted line
  indicates the average strength of the bump in the sample of
  \cite{Weedman 2006}, selected from the SWIRE survey.  The dashed
  curves indicate the expected locus for the mixture of a starburst
  and a quasar (see Section \ref{sec: bump-plot}). The $\tau=0$\ curve
  has no additional reddening beyond that in the starburst spectrum,
  while the $\tau=1$\ curve includes an addition reddening of
  $\tau(9.7\mu{\rm m})=1$.  Labeled tick marks indicate the 10\%\
  and 50\%\ AGN contribution to the mixture of normalized 1--1000
  \mic\ luminosities.  }
\end{figure}

\noindent This parameter is large when the slopes show an inflection, as expected
for a strong 1.6 \mic\ bump, but it is small when the slope of the SED
is constant.  In Figure \ref{fig: 6.2-bump}\ we show the strength of
the $B$\ parameter versus the equivalent width of the 6.2 \mic\ PAH
emission feature for the UDF sources compared to the local ULIRGs and
other high redshift targets.  The value of $B$\ was obtained by
interpolating the observed SED to get the rest-frame $JHK$\ 
photometry.  In the figure, we indicate the average strength of the
bump in a sample selected on the basis of that feature, by plotting
the average value of $B$\ for objects observed by \cite{Weedman 2006},
who obtained IRS spectra of sources in SWIRE fields with strong 1.6
\mic\ bumps.  Note that the plotted value is not a selection limit,
merely an indicative value.  The local sources used for comparison are
local ULIRGS observed by \cite{Armus 2006}.  The near-IR photometry is
taken from 2MASS \citep{Skrutskie 2006}, using the Large Galaxy Atlas
\citep{Jarrett 2003}\ where available and the Extended Objects catalog
otherwise.  We also plot the ratio that would be expected given the
NGC 7714 plus QSO model for composite objects.  We find that there is
a general trend of high PAH EQW for objects with large 1.6 \mic\ bumps.
While the uncertainties are large, the trend is about what is expected
from the composite spectrum.

A number of factors complicate Figure \ref{fig: 6.2-bump}.  First, the
model assumes that there is little extinction at 1.6 \mic.  Strong
dust reddening will reduce the value of $B$\ by lowering the $J$-band
flux relative to the $K$-band.  The effect of additional reddening is
shown in the figure.  Secondly, the strength of the bump depends on
the fraction of old stars in the source while the PAH emission depends
on the number of young stars. In a sense, the average age of the
population and the amount of dust determines an object's location in
Figure \ref{fig: 6.2-bump}. Thus a source with a much older underlying
population than NGC 7714 would have a higher ratio of $B$\ to EQW(6.2)
than the composite model.  Conversely, a very young object lacking
significant old stars could have a much lower ratio.  Finally, we have
implicitly assumed that the geometry of all sources is similar.  Both
the {\it Spitzer}\ photometry and spectroscopy of high redshift
sources capture the light of the entire source, because high redshift
sources are unresolved and fall within the entrance slit of the IRS.
For local ULIRGs, on the other hand, the IRS entrance slit usually
includes most of the MIR flux \citep{Soifer 2000, Armus 2006, Surace
  2007}, but is much smaller than the near-IR extent \citep{Scoville
  2000}.  Therefore, for the local ULIRGs in Fig. \ref{fig: 6.2-bump}\
we have used the total NIR light instead of an aperture matched to the
size of the IRS slit, simply to make comparison to the high-z sources
less problematic.

\subsection{Comparison with Other Surveys}

The set of high redshift IRS spectra considered here is small, but we examine if it is
typical of other objects targeted in this flux range.

First, we compare the properties of the faint, high-$z$\ starbursts with
more luminous high-z objects previously observed with the IRS.  The
largest samples were selected either using optical-to-MIR color
properties or by  selection at a longer wavelength, e.g.\ the
sub-millimeter \citep{Lutz 2005}.  \cite{Houck 2005}\ selected objects
based on very faint ($I_{AB}>25$) optical magnitudes and detectable
($\ge 0.8$\ mJy) 24 \mic\ flux densities.  \cite{Yan 2005}\ used a
less severe $R-[24]$\ selection function and also included the
requirement of blue 8 to 24 \mic\ color.  

The sources in the present sample are brighter in the optical than the
Houck et al.\ sources, being detected at $I_{AB}\sim 24.5$\ while the
latter were undetected at the same brightness.  The UDF sources are
much bluer, as the Houck et al.\ sources are 5-10 times brighter at 24
\mic. This difference is not surprising given the rarity of the Houck et al.\
sources (33 in 8 square degrees).  The $z>2$\ sources in the present
sample meet the Yan et al.\ selection criteria, but the lower redshift sources
are bluer in $R-24$\ color.  The Houck et al.\ and Yan et al.\ samples find
substantial numbers of objects with spectra showing little PAH
emission, likely indicating AGN-dominated sources.  These objects are
brighter than those in the current data by a factor of $\sim 5-7$,
suggesting bolometric luminosities in excess of $10^{13}$\ \lsun\ if
starburst-dominated templates are used; if the sources were
AGN-dominated, a smaller bolometric correction would be appropriate
and they might be closer in luminosity to the UDF sources.  Two
sources in the Yan et al.\ sample have measured PAH emission; the
EQW(6.2) in one object is consistent with the UDF sources while the
other has significantly weaker PAH emission.  

The GOODS-South region has not been observed at submm wavelengths, so
a direct comparison with SMGs is not possible.  \cite{Lutz 2005}\ 
obtained IRS spectra of two SMGs at $z\sim 2.8$, one of which shows
strong PAH emission while the other is dominated by a red AGN
continuum over the same rest-frame wavelength region covered by the
UDF spectra.  Lutz et al.\ place the AGN-dominated SMG in the same
region of the $L_X$/\lir\ plane identified by \cite{Alexander 2005}.
\cite{Menendez-Delmestre 2006}\ have also observed SMGs with the IRS.
They find the SMG spectra to be consistent with that of a composite of
a strongly star-forming galaxy and weak AGN.  The EQW of the 7.7 \mic\ 
PAH feature in their spectra is consistent with that of the $z>2$\ UDF
sources.

As discussed above, Source 1 at $z=2.69$\ provides a striking example
of an object that would be excluded by a ``bump'' survey but
nonetheless may have star formation contributing the majority of its
bolometric luminosity.  Sources of this kind may be an important
contribution to the cosmic star formation density.  However, based on
current data, it is difficult to estimate how many such sources will
be missed in IRAC surveys.  This uncertainty demonstrates the need for
wider use of deep MIR spectroscopy of high redshift sources.

\section{Summary}

We have obtained the longest exposure-time spectra to date using the
the {\it Spitzer}\ IRS, demonstrating the power of the IRS to trace
the properties of LIRGs/ULIRGs at $z>1$.  The spectra of four high
redshift sources in the {\it Hubble}\ UDF show prominent PAH emission,
comparable to local starbursts.  The combination of IRAC photometry,
IRS spectroscopy and X-ray data on these sources place constraints on
the physical properties which dominate their luminosities.  These
sources follow the rough correlation in ULIRGs between the strength of
PAH emission (arising from young stars) and the 1.6 \mic\ bump
(arising from older stars).

  \acknowledgements

This work is based in part on observations made with the {\it Spitzer
  Space Telescope}, which is operated by the Jet Propulsion
Laboratory, California Institute of Technology under NASA contract
1407. Support for this work was provided by NASA through an award
issued by JPL/Caltech.

\clearpage
\begin{deluxetable}{llllrrr}
\tablecaption{Photometry \label{Tab: phot}}
\rotate
\tablehead{
\colhead{Obj.} &
\colhead{R.A.} &
\colhead{Dec.} &
\colhead{$I_{AB}$} &
\colhead{$f_{8}$} &
\colhead{$f_{16}$} &
\colhead{$f_{24}$} \\
\colhead{} & 
\colhead{(J2000)} &
\colhead{(J2000)} &
\colhead{(mag.)} & 
\colhead{(mJy)} & 
\colhead{(mJy)} & 
\colhead{(mJy)} 
}

\startdata

1     & 3:32:44.00 & -27:46:35.0 & 24.47 & $0.0201 \pm 0.0004$ &  $0.05 \pm 0.025$ & $0.126 \pm 0.010$ \\  
1-BzK & 3:32:38.52 & -27:46:33.5 & 24.65 & $0.010 \pm 0.005$ &  $0.10 \pm 0.025$ & $0.200 \pm 0.020$ \\
2     & 3:32:34.85 & -27:46:40.4 & 23.30 & $0.0153 \pm 0.0004$ &  $0.15 \pm 0.025$ & $0.133 \pm 0.015$ \\
2-x   & 3:32:39.72 & -27:46:11.3 & 24.96 & $0.0467 \pm 0.0004$ &  $0.16 \pm 0.025$ & $0.225 \pm 0.020$

\enddata
\end{deluxetable}

\clearpage

\begin{deluxetable}{llllclcll}
\tablecaption{Observations \label{Tab: obs}}
\tablehead{
\colhead{Obj.} &
\colhead{$z$} &
\colhead{Sub-slit} &
\colhead{Mode} & 
\colhead{Ramp} &
\colhead{Cycles} &
\colhead{AORs} \\
\colhead{} & 
\colhead{} & 
\colhead{} & 
\colhead{} & 
\colhead{Time (s)} & 
\colhead{per AOR} & 
\colhead{} 
}

\startdata

1 & 2.69 & LL-1       & $6\times 1$\ map & 120 & 20 $\times$\ 6 pos.     & 3  \\ 
2 & 1.09 & SL-1       & Staring          & 240 & 25 $\times$\ 2 nods     & 1  \\
  &      & LL-2       & Staring          & 120 & 70 $\times$\ 2 nods     & 2

\enddata

\end{deluxetable}

\clearpage

\begin{deluxetable}{llccccccc}
\rotate
\tablewidth{8in}
\tablecaption{Emission Features \label{Tab: EW}}
\tablehead{
\colhead{Obj.} &
\colhead{$z$} &
\colhead{log(\lir/\lsun)\tablenotemark{a}} &
\colhead{Flux(6.2)\tablenotemark{b}} & 
\colhead{(EQW$_{\mbox{rest}}$)\tablenotemark{c}} &
\colhead{Flux(7.7)\tablenotemark{b}} &
\colhead{(EQW$_{\mbox{rest}}$)\tablenotemark{c}} & 
\colhead{Flux(8.6)\tablenotemark{b}} &
\colhead{(EQW$_{\mbox{rest}}$)\tablenotemark{c}} 
}

\startdata

 1     & 2.69      &  12.6 & $5.7\pm 0.7$ & ($0.24\pm0.06$)                 & $23.3\pm2.0$ & ($0.98\pm0.23$) & $5.7\pm0.7$ & ($0.39\pm0.09$)    \\
 1-BzK & 2.55      &  12.6 & $6.4\pm 0.5$  & ($0.24\pm0.09$)                 & $20.1\pm2.1$ & ($0.79\pm0.29$) & \nodata & \nodata \\
 2     & 1.09      &  11.5 & $15.3\pm1.0$ & ($0.6\pm0.2$)\tablenotemark{d} & $29.9\pm1.6$ & ($0.90\pm0.22$) & $5.8\pm1.6$ & ($0.22\pm0.07$) \\
 2-x   & $\sim1.55$\tablenotemark{e}  &  12.3 &  \nodata & \nodata & \nodata & \nodata & \nodata & \nodata 

\enddata
\tablenotetext{b}{The infrared luminosity inferred from the use of local templates as discussed in Section \ref{sec: newLIR}.  As only one local template
is used per object, the uncertainty is not given.  Typical uncertainties using a suite of templates is $<40$\%\ \citep{Chary and Elbaz 2001}.}
\tablenotetext{b}{$10^{-16}$\ ergs cm$^{-2}$\ s$^{-1}$}
\tablenotetext{c}{EW given in \mic.  In most cases, the uncertainty is dominated by the uncertainty in the continuum.}
\tablenotetext{d}{The continuum level is only marginally detected in the SL spectrum,
so this measurement is uncertain}.
\tablenotetext{e}{The redshift estimate is based on the identification of an 
isolated emission line as the 6.2 \mic\ PAH line.  The continuum is not well
detected, so EQW values are not given.}

\end{deluxetable}

\clearpage

\clearpage

\clearpage

\clearpage

\clearpage

\clearpage

\clearpage

\clearpage

  \clearpage

\clearpage

\end{document}